\documentclass[onecollarge,preprint]{svjour2}       
\usepackage[numbers,sort&compress]{natbib}
\smartqed  
\usepackage{graphicx}
\usepackage{amssymb}
\usepackage{amsmath}
\usepackage{stfloats}
\usepackage{dcolumn}
\usepackage{bm}
\usepackage{float}
\usepackage{caption}
\usepackage{lineno}
\usepackage[colorlinks,linkcolor=blue,anchorcolor=green,citecolor=blue,urlcolor=blue]{hyperref}
\hypersetup{colorlinks=true}

\usepackage{mathptmx}      

\journalname{Acta Mechanica [special issue for ACMFMS2016]}

\begin{document}

\title{Strain-mediated magnetoelectric effect for the electric-field control of magnetic states in nanomagnets}

\author{Min Yi         \and
        Bai-Xiang Xu   \and
        Dietmar Gross 
}


\institute{M. Yi, B.-X. Xu\at
Mechanics of Functional Materials Division, Technische Universit\"at Darmstadt, 64287 Darmstadt, Germany \\
              Tel.: +49 6151 16-22922\\
              Fax: +49 6151 16-21034\\
              \email{yi@mfm.tu-darmstadt.de; xu@mfm.tu-darmstadt.de}           
           \and
           D. Gross \at
Division of Solid Mechanics, Technische Universit\"at Darmstadt, 64287 Darmstadt, Germany \\
\email{gross@mechanik.tu-darmstadt.de}
}

\date{Received: December 27, 2016 / Accepted: date}

\maketitle

\begin{abstract}
Electric-field control of magnetism without electric currents potentially revolutionizes spintronics towards ultralow power. Here by using mechanically coupled phase field simulations, we computationally demonstrate the application of the strain-mediated magnetoelectric effect for the electric-field control of magnetic states in heterostructure. In the model heterostructure constituted of the soft nanomagnet Co and the piezoelectric substrate PMN-PT, both the volatility of magnetic states and the magnetization switching dynamics excited by the electric field are explored. It is found that an electric field can drive the single-domain nanomagnet into an equilibrium vortex state. The nanomagnet remains in the vortex state even after removing the electric field or applying a reverse electric field, i.e. the vortex state is extremely stable and nonvolatile. Only by utilizing the precessional magnetization dynamics, the 180$^\circ$ magnetization switching is possible in small-sized nanomagnets which are free of the stable vortex state. Electric-field pulses can realize the deterministic 180$^\circ$ switching if the electric-field magnitude, pulse width, and ramp time are carefully designed. The minimum switching time is found to be less than 10 ns. These results provide useful information for the design of low-power, reliable, and fast electric-field-controlled spintronics.
\end{abstract}

\section{Introduction}
\label{intro}
Magnetic systems are promising candidates for the development of fast, high-density, nonvolatile memory technology \cite{1}. Two typical examples of commercial storage systems are magnetic random access memory (MRAM) and spin transfer torque magnetic random access memory (STT-MRAM). Thereby, the magnetic state of nanomagnets is used for the information storage. The switching or writing process is driven by an external magnetic field in MRAM or by a spin-polarized electric current in STT-MRAM. In MRAM, built-in wires in every memory cell are required for the switching of a nanomagnet, i.e. the magnetic field must be generated by passing a current through a wire. The extra wires not only make the device circuit complicated and thus hinder the high density, but also generate current to result in energy dissipation and overheating. In contrast, by using the spin transfer torque effect, STT-MRAM allows for high density. But it is not a low-power method. Usually, a high current density in the order of $10^{10}-10^{11}$ A/m$^2$ \cite{2} is required for the switching of a nanomagnet with an acceptable endurance, leading to high energy dissipation. Therefore, in order to realize the low-power and high-density memory devices, a method without electric currents are ideally desired.

Recently, the control of magnetism by pure electric fields without electric currents provides one viable way for the future low-power next-generation spintronics based memory devices. This strategy can be realized in multiferroic materials which possess more than one ferroic effects and the coupling between two of them. For example, through the magnetoelectric (ME) coupling between the electrical polarization of a ferroelectric material and the magnetization of a ferromagnet, the control of magnetization by an electric field is achievable. Nevertheless, the ME coupling is weak in single phase systems. Such a control is usually implemented through the ME coupling in heterostructures. Generally, in ME heterostructures an electric field can manipulate the magnetization through the interfacial mechanisms such as strain-mediated elastic coupling \cite{3,4,5,6,7,8,9,10,11,12,13,14,15,16,17,18,19,20,21,22,23,24,a1,w1,v1,v2}, charge modulation \cite{25,26,27,28,29,30,31,32,33,34,v1,v2}, interface bonding \cite{35,36,37,38,39}, and exchanging coupling \cite{20,40,41,42,43}. Since strong elastic coupling between ferroelectric and ferromagnetic phases at room temperature can be achieved across their interfaces, strain-mediated ME effect is one of the most promising candidates and it is thus most extensively investigated in ferromagnetic/ferroelectric heterostructures. In this scenario, a strain generated in a ferroelectric layer by an electric field is transferred to the ferromagnetic layer through the interface and thus controls the magnetization through the magnetoelastic coupling. Due to the long-range nature of the elastic coupling, the strain-mediated ME effect works at the bulk level. On the contrary, all the other three mechanisms are localized to the region near the interface.

In order to establish the strain-mediated ME effect as a feasible strategy in the future practical memory devices driven by electric fields, two technical features should be clarified. Firstly, only if the ferromagnet persists in the new state and does not revert back to the old state after withdraw of the electric field, can this ME effect be applied to memory devices. It means that the magnetic states induced by the electric field should be nonvolatile. Secondly, an electric-field induced deterministic 180$^\circ$ magnetization switching should be achievable, since it determines the reliability and performance of spintronics based memory devices. For example, in order to attain a high signal-to-noise ratio in magnetic tunnel junction (MTJ) for spintronics, a significantly large electric resistance change of MTJ is required, which can only be achieved by a 180$^\circ$ switching in the free layer of MTJ. Trailblazing experiments have demonstrated strain-mediated \cite{14,34}, charge-mediated \cite{31}, and exchange-coupling-mediated \cite{42} voltage-driven 180$^\circ$ switching in Ni/BaTiO$_3$, BiFeO$_3$-based, and CoFeB/MgO/CoFeB heterostructures, respectively. From the theoretical side, a large number of studies have focused on the strain-mediated 180$^\circ$ switching either by designing the magnet shape \cite{21} or by using the 3D precessional switching dynamics \cite{4,6,9,5,15,10,18,16}.

In this work, we study both the volatility of magnetic states and the magnetization switching dynamics excited by the electric field through mechanically coupled phase field simulations. By taking the model heterostructure system which contains the soft nanomagnet Co and the piezoelectric substrate 0.7Pb(Mg$_{1/3}$Nb$_{2/3}$)O$_3$-0.3PbTiO$_3$ (PMN-PT), we analyze both the equilibrium magnetic states and the precessional dynamic switching process, in order to shed light on how to achieve nonvolatility and 180$^\circ$ switching.

\section{Model and simulations}
\begin{figure}
\centering
\includegraphics[width=8cm]{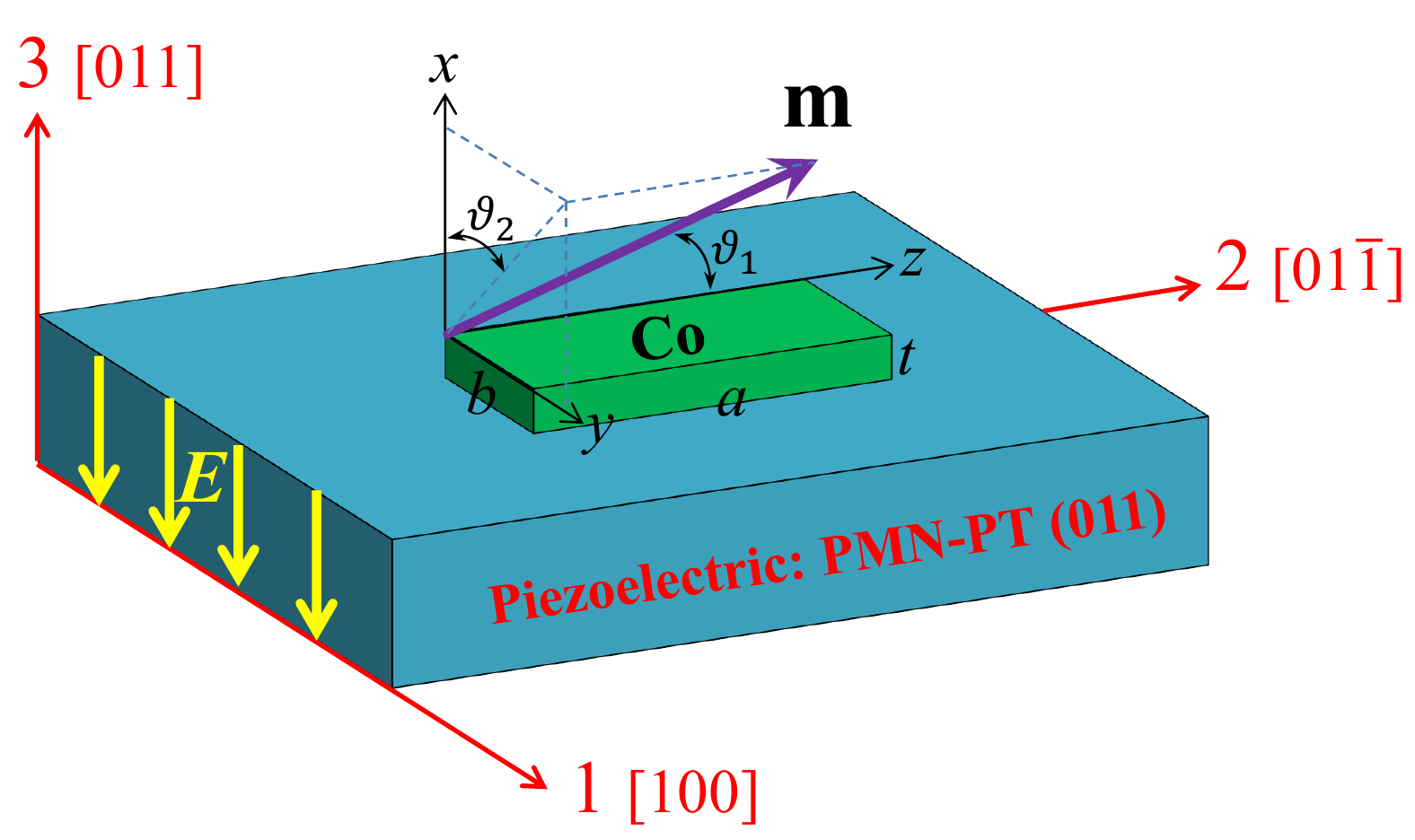}
\caption{ Model heterostructure containing the nanomagnet Co and the piezoelectric substrate PMN-PT (011).}
\label{fig1}
\end{figure}

The model heterostructure is shown in Fig. \ref{fig1}, including a nanomagnetic layer Co and a piezoelectric substrate PMN-PT. The nanomagnet is assumed to be much smaller than the substrate so that the strain transferred from the PMN-PT to the Co nanomagnet is nearly uniform. Meanwhile, due to the small magnetostrictive coefficient and small size of Co, the magnetization rotation or switching in Co has little influence on the electromechanical behavior of the piezoelectric substrate. PMN-PT with (011) orientation is used due to its large piezoelectric coefficients $d_{31}=−3175$ pC/N [100] and $d_{32}=1426$ pC/N [01$\bar{1}$] \cite{44}. The magnetic state evolution is simulated by a constraint-free phase field model \cite{45,46}. As shown in Fig. 1, the polar and azimuthal angles ($\vartheta_1$, $\vartheta_2$) instead of the Cartesian components ($m_1$, $m_2$, $m_3$) of the unit magnetization vector are taken as the order parameters. In this way, the constraint of constant magnetization magnitude at the temperature far below 
the Curie point is fulfilled automatically. The total magnetic enthalpy for the nanomagnet includes the magnetocrystalline anisotropy contribution $\mathcal{H}^\text{ani}$ , the exchange contribution $\mathcal{H}^\text{exc}$, the magnetostatic contribution $\mathcal{H}^\text{mag}$, the pure mechanical contribution $\mathcal{H}^\text{mech}$, and the magneto-elastic coupling contribution $\mathcal{H}^\text{mag-ela}$, i.e.
\begin{equation}
 \begin{split}
  \mathcal{H} =\mathcal{H}^\text{ani} + \mathcal{H}^\text{exc} + \mathcal{H}^\text{mag} + \mathcal{H}^\text{mech} + \mathcal{H}^\text{mag-ela}
 \end{split}
 \label{mag-enthalpy}
\end{equation}
in which
\begin{equation}
\begin{split}
 \mathcal{H}^\text{ani} =& K_u \sin^2\vartheta_1 \\
 \mathcal{H}^\text{exc} =& A_e (\vartheta_{1,j}\vartheta_{1,j} + \sin^2\vartheta_1 \vartheta_{2,j}\vartheta_{2,j}) \\
 \mathcal{H}^\text{mag} =& -\frac{1}{2}\mu_0H_jH_j - \mu_0 M_s (H_1\sin\vartheta_1\cos\vartheta_2 
 + H_2\sin\vartheta_1\sin\vartheta_2+H_3\cos\vartheta_1) \\
    \mathcal{H}^\text{mech} =&
  \frac{1}{2}C_{11}(\varepsilon_{11}^2+\varepsilon_{22}^2) + \frac{1}{2}C_{33}\varepsilon_{33}^2 
   + C_{13}(\varepsilon_{11}\varepsilon_{33}+\varepsilon_{22}\varepsilon_{33}) \\
  & + C_{12}\varepsilon_{11} \varepsilon_{22} + 2C_{44}(\varepsilon_{23}^2+\varepsilon_{31}^2) 
   + (C_{11}-C_{12})\varepsilon_{12}^2 \\
  \mathcal{H}^\text{mag-ela} =
 & B_1 (\sin^2\vartheta_1\cos^2\vartheta_2\varepsilon_{11} 
      + 2\sin^2\vartheta_1\sin2\vartheta_2 \varepsilon_{12} 
  + \sin^2\vartheta_1\sin^2\vartheta_2 \varepsilon_{22}) \\ 
 &  + B_2 \sin^2\vartheta_1\varepsilon_{33} + B_3 \sin^2\vartheta_1(\varepsilon_{11}+\varepsilon_{22}) \\
 &  + B_4 (\sin2\vartheta_1\sin\vartheta_2 \varepsilon_{23} + \sin2\vartheta_1\cos\vartheta_2 \varepsilon_{31}) 
\end{split}
\label{ela-enth}
\end{equation}
Here, $C_{11}$, $C_{12}$, $C_{33}$, $C_{13}$, and $C_{44}$ are the elastic constants.
$\varepsilon_{ij}=\frac{1}{2} \left(u_{i,j}+u_{j,i} \right)$ are the strain components with $u_i$ as the mechanical displacement. The Latin indices $i$ and $j$ run over 1--3. $B_1$, $B_2$, $B_3$, and $B_4$ are the magneto-elastic coupling coefficients. $K_u$ is the anisotropy constant. $M_s$ is the saturation magnetization. $A_e$ is the exchange stiffness constant. $\mu_0$ is the vacuum permeability. The magnetic field is expressed as the negative gradient of the magnetic scalar potential $\phi$, i.e. $H_j=-\phi_{,j}$.

Based on the above magnetic enthalpy, a combination of the balance law of the configurational force system and the second law of thermodynamics can give rise to a generalized evolution equation for the order parameters $\vartheta_1$ and $\vartheta_2$, which takes the form \cite{45,46}:
\begin{equation}
 \frac{1}{M_s}\left(\frac{\partial \mathcal{H}}{\partial \vartheta_{\mu,j}}\right)_{,j} 
  - \frac{1}{M_s}\frac{\partial \mathcal{H}}{\partial \vartheta_\mu} + \zeta_\mu^\text{ex} 
  = \frac{1}{\gamma_0}L_{\mu\gamma}\frac{\partial \vartheta_\gamma}{\partial t}
 \label{evolution}
\end{equation}
in which
\begin{equation}
 L_{\mu \gamma}= \left[\begin{array}{cc} \alpha & -\sin\vartheta_1 \\ \sin\vartheta_1 & \alpha \sin^2\vartheta_1 \end{array} \right],
\end{equation}
$\zeta_\mu^\text{ex}$ is the external driving force for $\vartheta_1$ and $\vartheta_2$, $\alpha$ the damping coefficient, and $\gamma_0=1.76\times10^{11}/(Ts)$ the gyromagnetic ratio. The Greek indices $\mu$ and $\gamma$ run over 1--2.

In addition, the mechanical equilibrium equation and the Maxwell equation which governs the magnetic part are incorporated:
\begin{equation}
 \left(\frac{\partial \mathcal{H}}{\partial \varepsilon_{ij}}\right)_{,j}=\bm{0}
 \qquad \text{and} \qquad
 \left(-\frac{\partial \mathcal{H}}{\partial H_j}\right)_{,j}=0
 \label{mechmageq}
\end{equation}

By using the six degrees of freedom $[u_1,u_2,u_3,\phi,\vartheta_1,\vartheta_2]^\text{T}$, a three-dimensional nonlinear finite element implementation in FEAP \cite{47} is performed to solve Eqs. (\ref{evolution}) and (\ref{mechmageq}). The nanomagnet size is $a=2b$ and $t=2$ nm. The materials parameters are taken as \cite{4,6,9}: $C_{11}$=307 GPa, $C_{12}$=165 GPa, $C_{44}$=75.5 GPa, $C_{13}$=103 GPa, $C_{33}$=358 GPa, $B_1$=$-$8.1 MPa, $B_2$=$-$29 MPa, $B_3$=28.2 MPa, $B_4$=29.4 MPa, $K_u$=6.5$\times10^4$ J/m$^3$, $M_s$=1.424$\times10^6$ A/m, $A_e$=3.3$\times10^{-11}$ J/m, $\alpha$=0.01. Due to the exchange length $\sqrt{2A_e/(\mu_0M_s^2)}\!\approx\!5.1$ nm, the finite element mesh size is chosen to be 2 nm. The strain generated by applying an electric field to the PMN-PT is estimated as $\varepsilon_{yy}=Ed_{31}$ and $\varepsilon_{zz}$=$Ed_{32}$ in the nanomagnet Co, as shown in Fig. \ref{fig1}

\section{Results and discussions}
\subsection{Electric field induced nonvolatile vortex states}
From the viewpoint of information storage such as binary logic and memory applications, the magnetic states triggered by an electric field should be nonvolatile. This requires that the magnetic states should be robustly kept after the removal of the electric field. As the first step, we explore the electric field induced equilibrium magnetic states, in order to figure out whether the nonvolatile state is achievable when the electric field is applied for a sufficient long time without a precise control of time. The initial state is a single domain along the $z$ axis (Fig. \ref{fig1}). Then an electric field is applied to drive the initial state into another magnetic state. Fig. 2 presents the equilibrium magnetic states when an electric field of $E=7$ kV/cm is applied. It can be found that the electric field induced equilibrium magnetic states are dependent on the the nanomagnet size. If the size is small (e.g. $a=24$ nm), a single-domain state survives with the magnetization perpendicular to the plane. Such a state possesses high magnetostatic energy and magnetocrystalline anisotropy energy, but low magneto-elastic energy. The role of the latter energy surpasses that of the former two energies in the small-sized sample, thus leading to a single domain. On the contrary, in the case of large size (e.g. $a=52$ nm), we attain a vortex structure in which the magnetization near the vortex core gradually tilts out of plane. The vortex is a result of substantially reducing the exchange energy and slightly increasing the magnetostatic energy and the magnetocrystalline anisotropy energy. In the region outside the vortex, the magnetization almost lies in the plane due to the dominating role of the magnetostatic energy. When the in-plane size is large, the magnetostatic energy will be very large if the magnetization deviates from the plane. If the in-plane size is neither too small nor too large (e.g. $a=32$ nm), the magnetostatic energy and the magneto-elastic energy are comparable. Then we have an intermediate state, in which neither single domain nor perfect vortex state exists. All the magnetization tilts out of the plane and only presents the trend of forming a vortex state. These results provide fruitful insight onto the electric-field control of magnetic states by tuning the nanomagnet size.

\begin{figure}
\centering
\includegraphics[width=13cm]{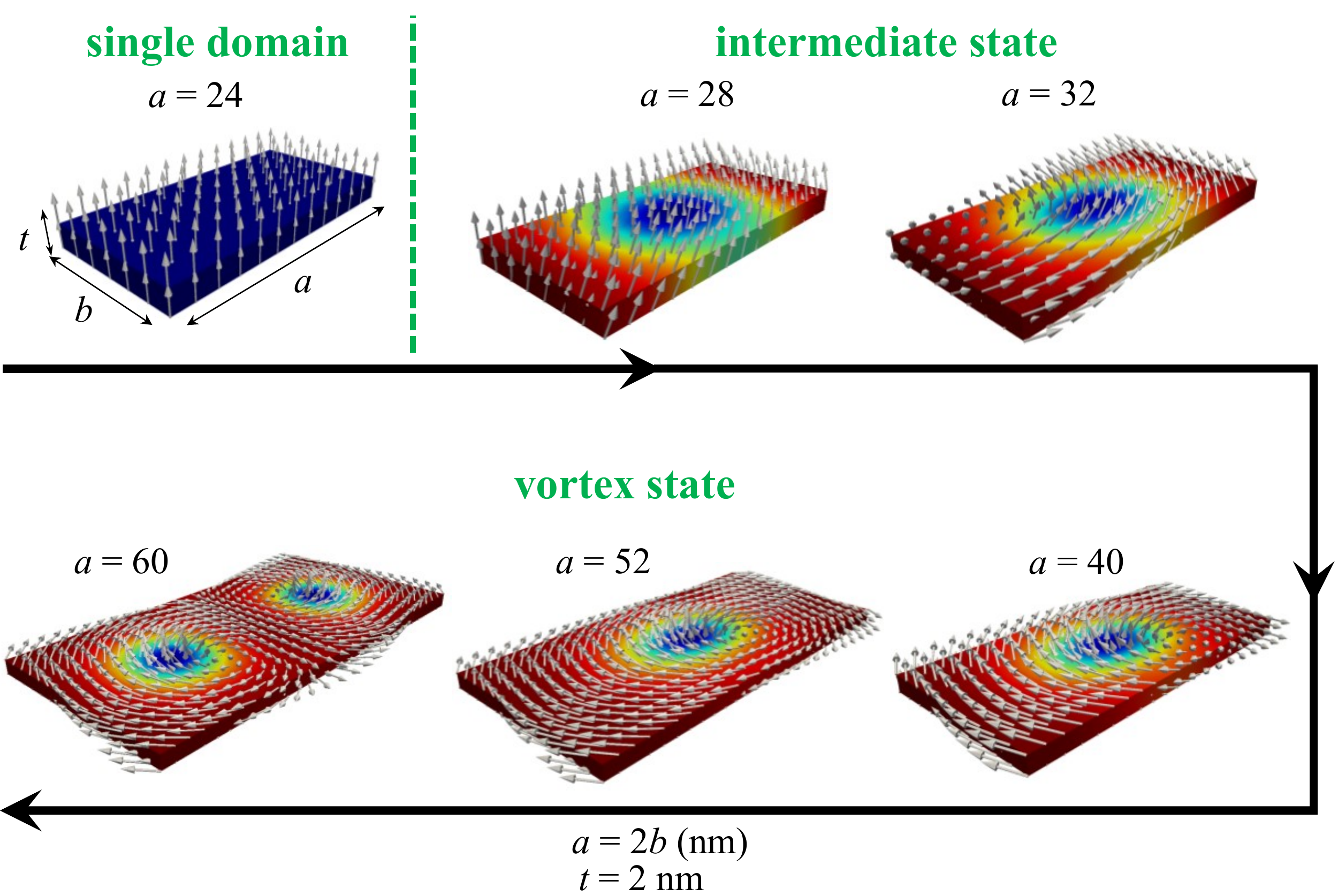}
\caption{Equilibrium magnetic states varying with the nanomagnet size in the case of $E=7$ kV/cm.}
\label{fig2}       
\end{figure}

\begin{figure}
\centering
\includegraphics[width=14cm]{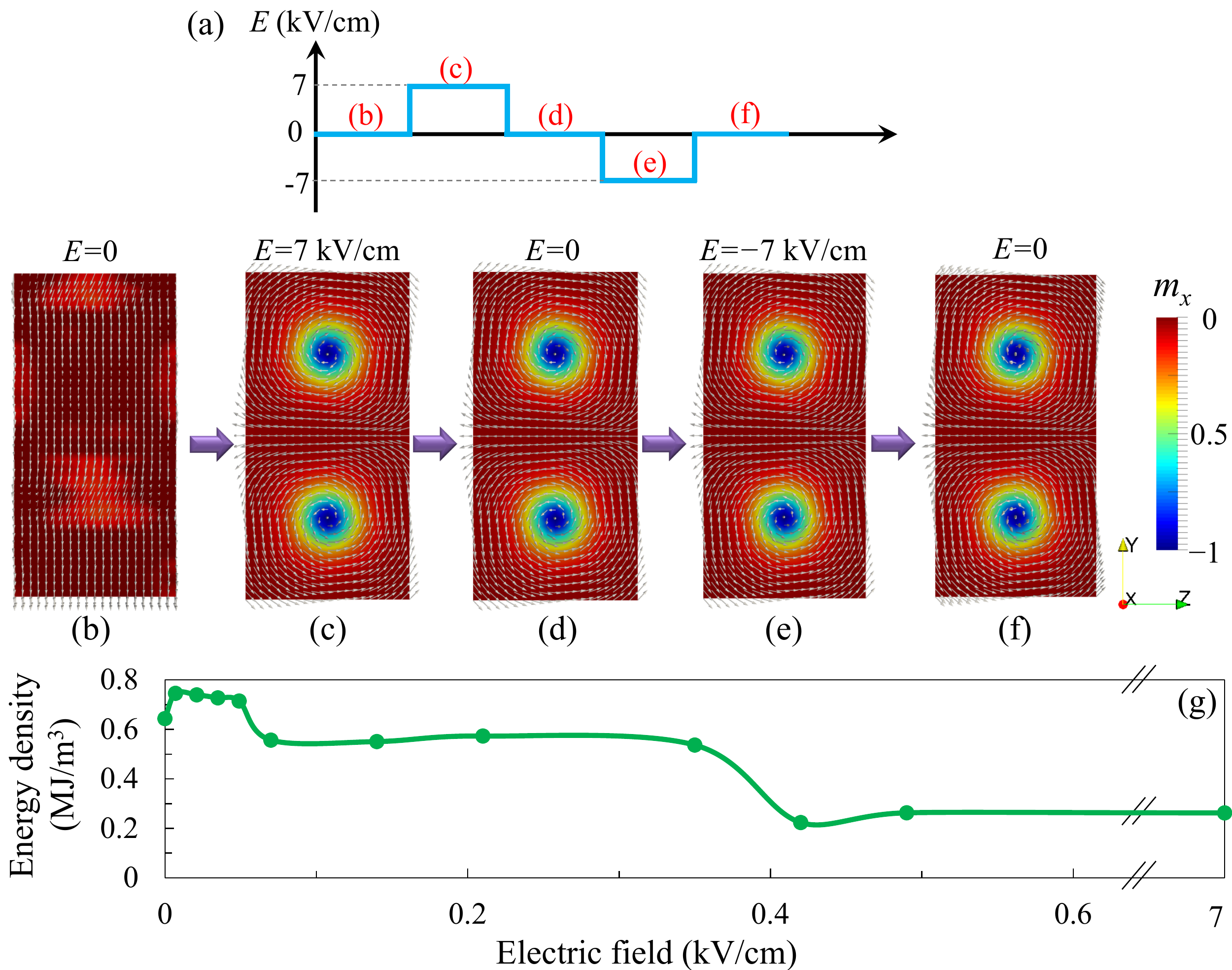}
\caption{The evolution of equilibrium magnetic states (b)-(f) in a nanomagnet when the electric field history in (a) is applied. (g) Total energy related to magnetization as a function of electric field. $a=2b=80$ nm, $t=2$ nm.}
\label{fig3}       
\end{figure}

It should be noted that the magnetic states shown in Fig. \ref{fig2} are achieved when a constant electric field is always kept. The stability of these magnetic states should be further examined. For example, in the case of equilibrium single domain (e.g. $a=24$ nm), we find that after removing the electric field, the magnetization will either revert back to the initial state or switch to the negative z axis. It indicates a volatile state. While the intermediate and vortex states can remain after the removal of the electric field and show nonvolatility. Fig. \ref{fig3}(b)-(f) presents the typical vortex state evolution in the case of an alternating electric field shown in Fig. \ref{fig3}(a). For every stage in Fig. \ref{fig3}(a), the electric field duration is long enough to make the magnetization reach the equilibrium state. Based on the initial single-domain state in Fig. \ref{fig3}(b), we find that the magnetic state transforms to the vortex state in Fig. \ref{fig3}(c) by applying an electric field of 7 kV/cm. This vortex state remains as the electric field vanishes in Fig. \ref{fig3}(d). A negative electric field of $-7$ kV/cm is subsequently applied, but it does not push the nanomagnet out of the vortex state, as shown in Fig. \ref{fig3}(e). Eventually, after removing the negative electric field again, the vortex state continues to remain (Fig. \ref{fig3}(f)). These features clearly demonstrate the nonvolatile vortex state. A similar vortex state, driven by acoustic waves, was also observed in elliptical cobalt nanomagnets in recent experiments \cite{48}.

Both the vortex state and the initial single-domain state are local energy minima. But in order to get the vortex state from the initial single-domain state, the nanomagnet has to experience the state with high exchange energy so that the formation of a curling magnetic structure is possible. Therefore, an energy barrier exists between these two states. When this energy barrier is overcome by the magneto-elastic energy which is introduced from the electric field, the magnetization will be driven from the initial single-domain state into the vortex state. Moreover, due to its local minimum, the vortex state can persist after the removal of the electric field. This explains the transition from Fig. \ref{fig3}(b) to Fig. \ref{fig3}(d). In addition, since the vortex state is extremely stable, the electric field is obviously not sufficient to drive the magnetization out of the vortex state. This can be verified by Fig. \ref{fig3}(c)-(f), in which the vortex state presents little difference no matter the electric field is 7, 0, or $-7$ kV/cm. Only a magnetic field is possible to restore the initial single-domain state in Fig. \ref{fig3}(b).
The evolution of energy density (related to magnetization) in Fig. \ref{fig3}(g) shows that a small electric field is required to drive the initial single-domain state into a magnetic curling state, along with an increase in the energy density. Then an electric field around 0.5 kV/cm is enough to stimulate the stable vortex state. The stable and nonvolatile vortex state paves an alternative way for designing memory devices.

\begin{figure}
\centering
\includegraphics[width=13cm]{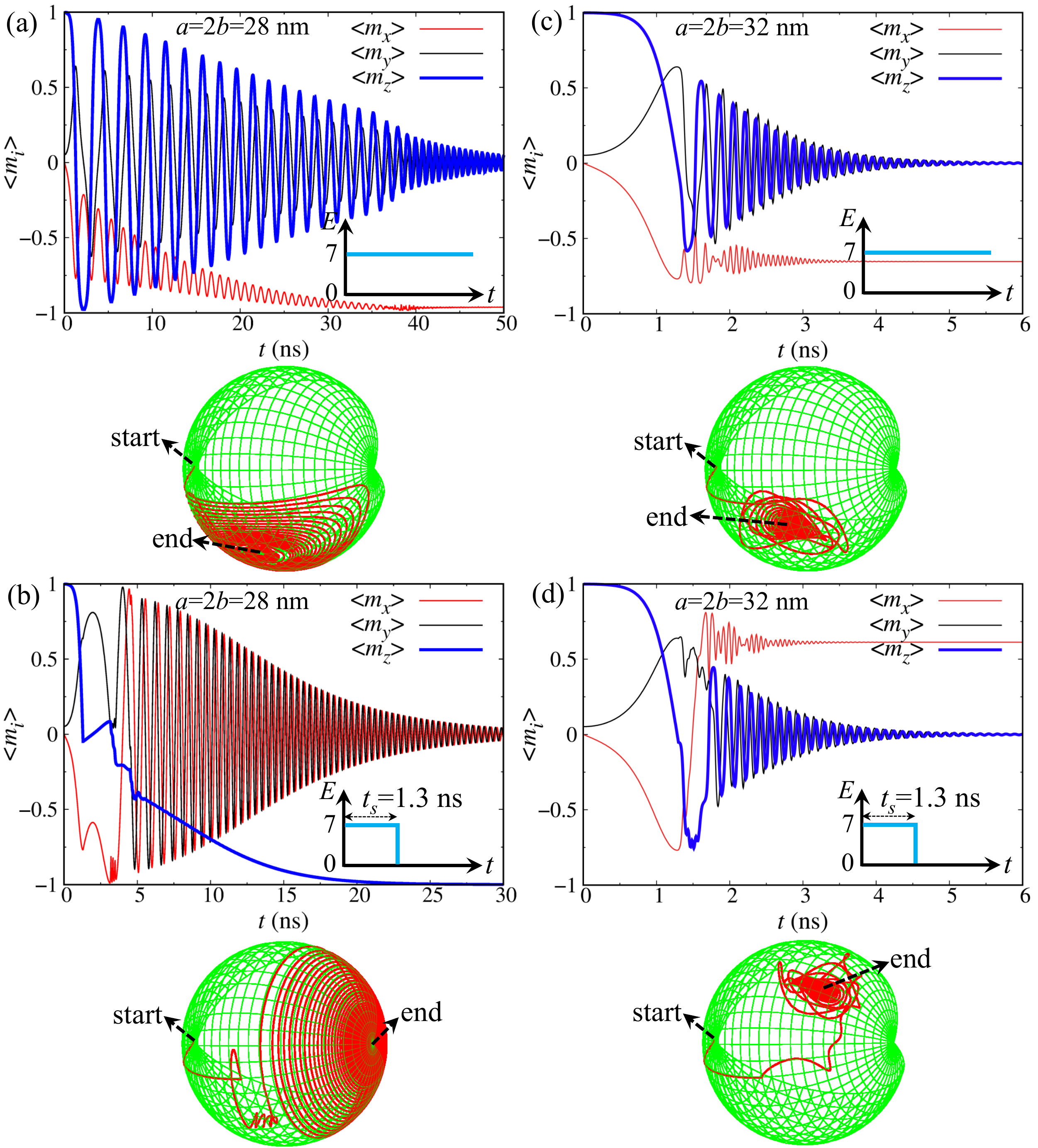}
\caption{Dynamics of average magnetization components $<m_i>$. An electric field of 7 kV/cm is always kept for the nanomagnet with (a) $a=2b=28$ nm and (c) $a=2b=32$ nm. An electric field of 7 kV/cm with a pulse width of 1.3 ns is applied for the nanomagnet with (b) $a=2b=28$ nm and (d) $a=2b=32$ nm.}
\label{fig4}
\end{figure}

\subsection{Electric field induced 180$^\circ$ switching}

\begin{figure}
\centering
\includegraphics[width=8cm]{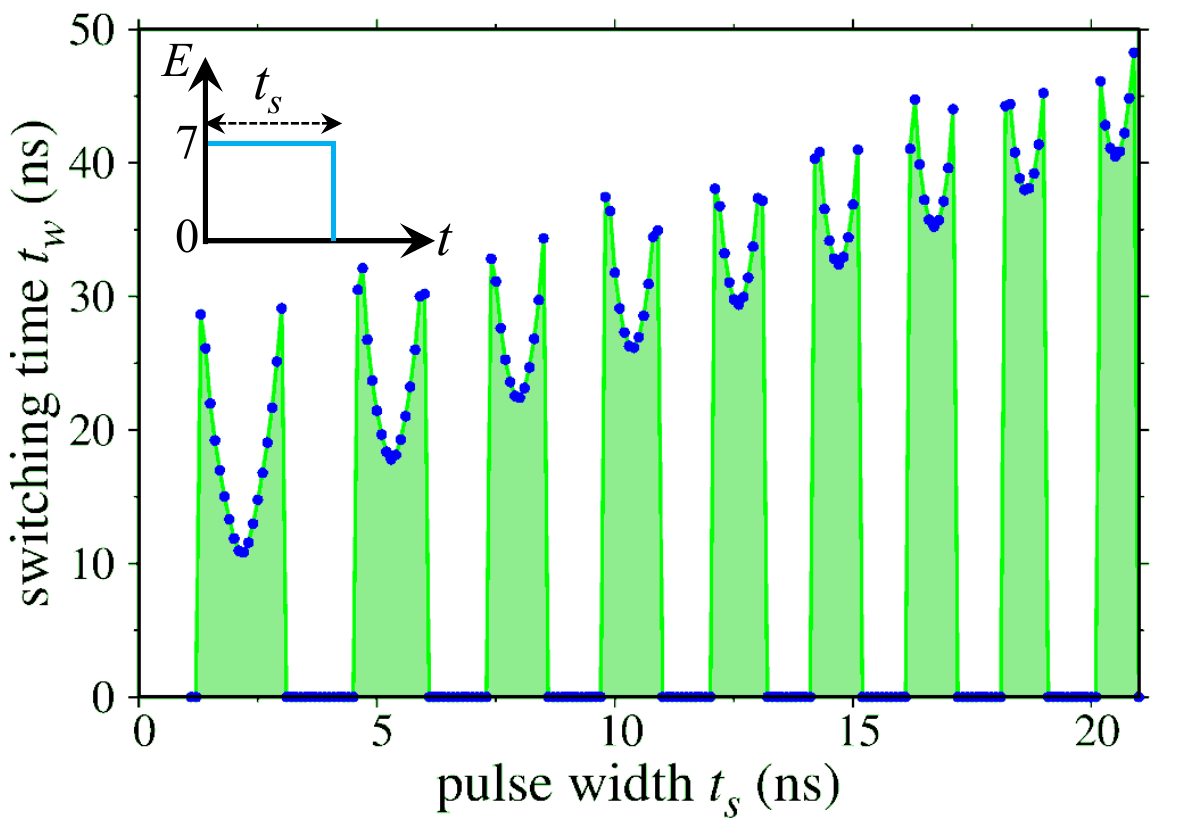}
\caption{180$^\circ$ switching time $t_w$ as a function of pulse width $t_s$ ($E=7$ kV/cm, $t_r=0$). In the regions between the green columns a 180$^\circ$ switching is unachievable.}
\label{fig5}
\end{figure}

\begin{figure}
\centering
\includegraphics[width=15cm]{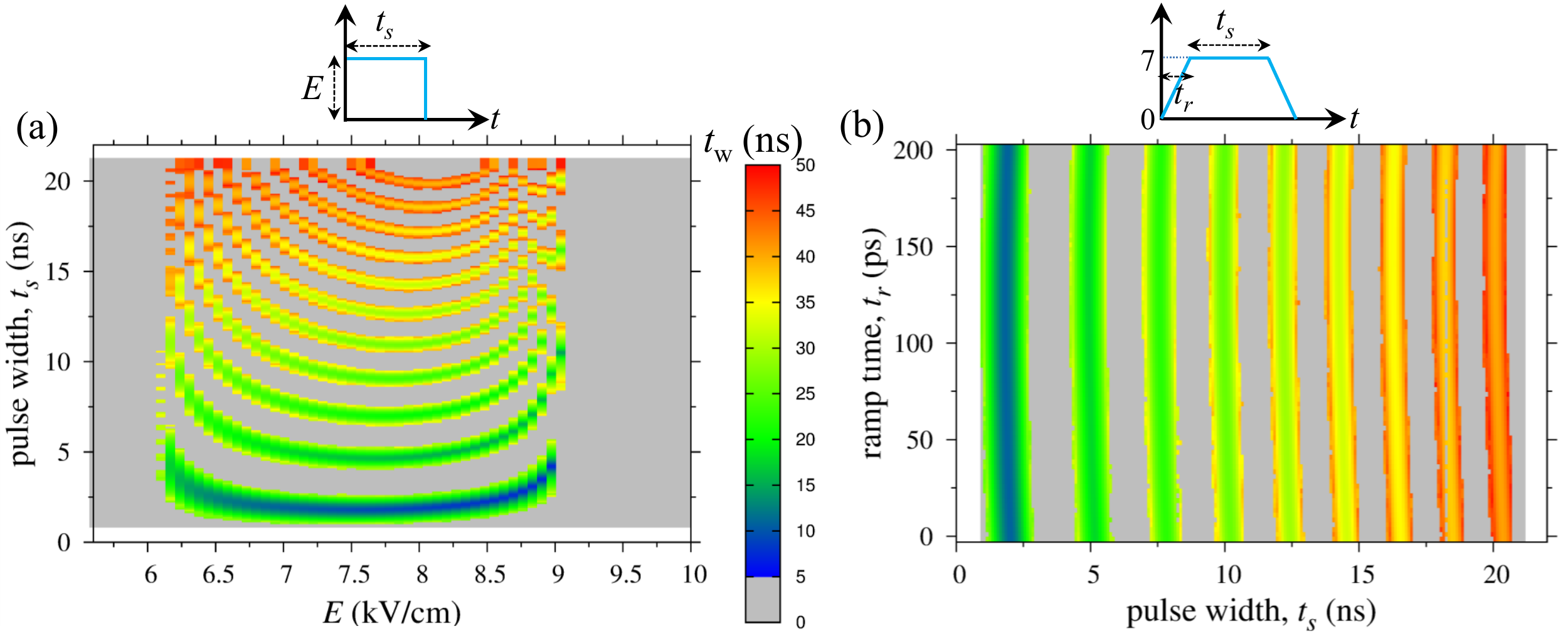}
\caption{180$^\circ$ switching time $t_w$ (a) as a function of electric field magnitude $E$ and pulse width $t_s$ ($t_r=0$), and (b) as a function of pulse width $t_s$ and ramp time $t_r$. The gray region means 180$^\circ$ switching unachievable.}
\label{fig6}
\end{figure}

The above study, focusing on the equilibrium magnetic states, implies that none of the nanomagnets experience a deterministic complete reversal of magnetization, i.e. 180$^\circ$ switching which also plays an important role in the spintronics based memory devices. How to achieve a 180$^\circ$ switching purely by an electric field is critical for the design of low-power spintronics. In contrast to the above study, here we apply an electric field pulse to trigger the 180$^\circ$ switching by utilizing the precessional magnetization dynamics. Fig. \ref{fig4} presents the typical results for the temporal evolution of average magnetization components for two nanomagnets with different size. It can be found in Fig. \ref{fig4}(a) and (c) that for both nanomagnets, new magnetic states similar to those in Fig. \ref{fig2} are formed if the electric field is always kept. No 180$^\circ$ switching occurs. However, if we apply an electric field pulse, the 180$^\circ$ switching is possible, as shown in Fig. \ref{fig4}(b) for the nanomagnet with a size of $a=2b=28$ nm. But if the nanomagnet size is increased to $a=2b=32$ nm, even an electric field pulse cannot achieve a 180$^\circ$ switching, as shown in Fig. \ref{fig4}(d). Such results indicate that the achievement of 180$^\circ$ switching relies on the nanomagnet size. Larger size implies a magnetization state more close to a vortex-like curling structure which is more stable (Fig. \ref{fig2}), thus prohibiting the 180$^\circ$ switching. Only in the case of relatively small size where magnetization is close to coherent rotation, a 180$^\circ$ switching is achievable if the precessional dynamics is utilized. In order to confirm the above size dependence, we perform a series of simulations for the size not smaller than $a=2b=32$ nm. It is found that, no matter how to control the electric field magnitude and pulse width, a 180$^\circ$ switching is seldom obtainable. It is absolutely impossible to achieve the 180$^\circ$ switching in the size range ($a>\sim 40$ nm) for the equilibrium vortex state in Fig. \ref{fig2}. In the case of size range ($a<\sim 24$ nm) for the single-domain state in Fig. \ref{fig2}, exactly coherent switching occurs and a 180$^\circ$ switching can be definitely achieved by controlling the electric field pulse. In the case of size range (24 nm $<a<$40 nm) for the intermediate state in Fig. \ref{fig2}, a 180$^\circ$ switching is possible for the size close to the single-domain region but impossible for the size close the vortex region.

After confirming the feasibility of achieving 180$^\circ$ switching through the precessional magnetization dynamics, we further study the effect of electric field magnitude, pulse width ($t_s$), and ramp time ($t_r$) on the switching behavior in the nanomagnet with $a=2b=28$ nm. Fig. \ref{fig5}(a) presents the switching time ($t_w$) as a function of $t_s$ in the case of $E=7$ kV/cm and $t_r=0$. It exhibits intermittent pulse-width regions (filled by green color) for the 180$^\circ$ switching events, inheriting the nature of precessional dynamics and oscillated behavior (e.g. Fig. \ref{fig4}(a) and (b)). In each intermittent region, a local minimum switching time can be determined, which increases with the pulse width. In the case of $t_s=2.2$ ns, a global minimum switching time of $\sim10$ ns can be realized. These results inspire a careful design of pulse width to develop a fast switching and fast memory technology.

We further manipulate all the electric field magnitude, pulse width, and ramp time to give a systematic insight onto the 180$^\circ$ switching condition and the switching time, as shown in Fig. \ref{fig6}. It can be found from Fig. \ref{fig6}(a) that the minimum electric field for a 180$^\circ$ switching is around 6.1 kV/cm and the maximum is around 9 kV/cm. In order to achieve a 180$^\circ$ switching, both the electric field magnitude and pulse width should be carefully designed. In the case of $E\sim 8.9$ kV/cm and $t_s=3.7$ ns, a fast switching with $t_w\sim 7.4$ ns can be realized. We also consider that the electric field is not applied instantly, i.e. a ramp time is required to increase $E$ from 0 to the final magnitude, as shown in Fig. \ref{fig6}(b). It is found that the ramp time $t_r$ within 200 ps has no strong influence on the switching time. For all the cases in Fig. \ref{fig6}, a 180$^\circ$ switching within less than 10 ns is attainable. It should be noted that the switching time of $<10$ ns here is close to that in the traditional STT-MRAM, MRAM, and DRAM (dynamic random access memory) \cite{49}, but without any electric currents.

\section{Conclusions}
The electric-field control of magnetic states in nanomagnets by the strain-mediated magnetoelectric effect in nanomagnetic/piezoelectric heterostructure has been studied by phase field simulations. It is found that depending on the nanomagnet size, the electric field induced equilibrium magnetic states can be either nonvolatile or volatile. For large nanomagnet size, a nonvolatile vortex states can be achieved, which persists after removing the electric field or applying a reverse electric field. For small nanomagnet size, coherent switching occurs and the electric field induced single-domain state is unstable and volatile. For the nanomagnet size between the above two cases, an intermediate magnetic state between single domain and vortex exists. By using the precessional magnetization dynamics in the case of single-domain state and intermediate state which is close to the single-domain state, a 180$^\circ$ switching can be achieved by an electric field pulse. On the contrary, for the case of vortex state and intermediate state which is close to the vortex state, a 180$^\circ$ switching is impossible. Careful design of the electric field magnitude, pulse width, and ramp time can result in a 180$^\circ$ switching time of less than 10 ns, which is close to that in the traditional STT-MRAM, MRAM, and DRAM. It is anticipated that the present study provides valuable insight into the design of electric-field control of nonvolatile magnetic states and 180$^\circ$ switching without electric currents for achieving low-power, high-speed, nonvolatile, and highly compact memory devices.

\begin{acknowledgements}
The financial supports from the German federal state of Hessen through its excellence programme LOEWE RESPONSE and the German Science Foundation (individual project Xu 121/7-1 and the project Xu 121/4-2 in the Forschergruppe FOR1509) are appreciated. The authors also greatly acknowledge the access to the Lichtenberg High Performance Computer of TU Darmstadt.
\end{acknowledgements}



\end{document}